\documentclass[aps,preprint]{revtex4}
\usepackage{epsf}

\begin{document}

\title{New Global Minima for Thomson's Problem of Charges on a Sphere }

\author{Eric Lewin Altschuler} 

\affiliation{Mt. Sinai School of Medicine
1425 Madison Avenue, Box 1240
New York, NY 10029\\
email: Eric.Altschuler@mssm.edu
}

\author{Antonio P\'erez--Garrido }
\affiliation{Departamento de F\'\i sica Aplicada
Universidad Polit\'ecnica de Cartagena, Campus Muralla del Mar
Cartagena, 30202
Murcia, Spain\\
email:Antonio.Perez@upct.es}

\begin{abstract}
Using numerical  arguments we find that for $N$ = 306 a tetrahedral configuration ($T_h$) and for $N=542$ a dihedral configuration ($D_5$) are likely the global energy minimum for Thomson's problem of minimizing the energy of $N$ unit charges on the surface of a unit conducting sphere.  These would be the largest $N$ by far, outside of the icosadeltahedral series, for which a global minimum for Thomson's problem is known. 
We also note that the current theoretical understanding of Thomson's problem does not rule out a symmetric configuration as the global minima for $N=306$  and 542.
 We explicitly find that analogues of the tetrahedral and dihedral configurations for $N$ larger than 306 and 542, respectively, are not global minima, thus helping to confirm the theory of Dodgson and Moore (Phys. Rev. B 55, 3816 (1997))  that as $N$ grows dislocation defects can lower the lattice strain of symmetric configurations and concomitantly the energy.  As well,
 making explicit previous work by ourselves and others,  
 for $N<1000$ we give a full accounting of icosadeltahedral configuration which are not global minima and those which appear to be, and discuss how this listing and our results for the tetahedral and dihedral configurations  may be used to refine theoretical understanding of Thomson's problem. 
\end{abstract}
\maketitle

What configuration of $N$ unit charges on the (surface) of a unit conducting sphere 
minimizes the Coulombic energy $\sum_{i\ne j}1/r_{ij}$ \cite{1}?  Beyond physics this
 question has utility in assembly of biological \cite{2} and chemical structures \cite{3,4}, to 
 benchmark optimization algorithms, and, as well, in mathematics Smale \cite{5} has 
 noted this question to be a {\it Hilbert problem} for the Twenty-First Century.  For $2\le 
 N\le 100$, the original question asked by J.J. Thomson a century ago, and a bit 
 beyond, there is agreement of all numerical \cite{6,7,8,9,10} and theoretical \cite{11} 
 methods suggesting that the global minimum configuration has been found.  However, 
 for larger $N$ owing to an exponential growth in good local minima \cite{7}, finding 
 global minima, general principles or insights for minimization, or even methods or cases 
 to test hypotheses has proven extremely difficult.   For $N = 10(h^2 + k^2 + hk) + 2$ 
 highly symmetric icosadeltahedral configurations can be constructed. While it was 
 initially thought that such configurations might be global minima \cite{12}, for large $N$ 
 adding defects to the icosadeltahedral lattice lowers the energy \cite{13,14,15,Toomre,Bowick}.  
 Here we note a tetrahedral configuration for $N = 306$ and a dihedral configuration
 for $N=542$ which based on numerical 
  arguments appear to be global minima, the largest such 
 $N$ by far, aside from the icosadeltahedral series, for which a global energy minimum 
 configuration is known.  Study of tetrahedral analogues larger than 306 and dihedral 
 analogues larger than 542 help confirm 
 the theory \cite{16} on defects lowering lattice strain and energy.  As well, we note that 
 the fact that lattice configurations fail to be global minima for $N > 800$, may help 
 explain why Mo$_{154}$ anions self-assemble into a spherical superstructure with a 
 non-lattice number of subunits \cite{4}.

For $N = 78$, as originally suggested by Edmunson \cite{11}, the presumed global 
minimum configuration has tetrahedral ($T_h$) symmetry (Figure 1$a$) \cite{7,8,9,10,11}.
We had previously suggested that an analogue of this configuration with 306 charges 
(Figure 1$b$, see below for method of construction of this analogue) also might be a global 
minimum and done limited numerical testing of this idea \cite{12}. Now, here we have 
extensively checked numerically on over one thousand runs, and have found no 
configuration of lower energy. 
(Numerically to look for non-lattice configurations with energies lower than the lattice
configuration we used random initial configurations followed by a local steepest 
descent method.  For $N = 306$ we could not find a configuration in one thousand
 runs with an energy lower than the lattice energy, though for many other $N$ 
considered in this paper 
including $N =$ 1218 and 4866, and smaller $N$ with icosadeltahedral lattices
 (see Tables I and II, and Figure 2) configurations with energies lower than the lattice 
 energy could be found in only fifteen runs.)
But one thousand runs is orders of magnitude less than 
the estimated \cite{7} over 1.5 million local minima for $N = 306$.  As well, without an 
analytic proof we could not be certain the tetrahedral configuration in Figure 1$b$ is the 
one of minimum energy for $N = 306$.
Current theoretical understanding of Thomson's problem \cite{13,Toomre,Bowick,16,17},
to be discussed below, does not exclude the possibility of a symmetric configuration
as the global energy minimum for  $N=306$.
Of course, our configuration stands  open to challenge.

Using a genetic algorithm Morris, Deaven and Ho \cite{9} confirmed previously found
\cite{10} presumed global minima for $N\le 112$ and gave their likely global minima for 
$N\le 200$. Fitting the energy of these minima for $N\le 200$ to the function:
$E=N^2/2\left( 1+aN^{-1/2}+bN^{-3/2}  \right)$, 
(see \cite{9} and refs.\ therein for an explanation of why this function was used)
for $N > $100 they found particularly deep minima with respect to this function for icosadeltahedral configurations for $N =$ 122, 132 192 and for dihedral $D_5$ configurations for $N =$ 137, 182 and 187 and $D_2$ configuration for N = 146.  
By our numerical testing as well these dihedral configurations appear to be global minima, though we have no analytic proof and numerically we are orders of magnitude short [7] to even sample a majority of local minima. We next looked at the
higher split analogues of these dihedral configurations with $4N- 6$ charges. 
(See below for the method of construction of such analogues.) The next larger dihedral analogues for $N =$ 146, 182 and 187--$N =$ 578, 722, and 742 respectively--are found not to be global minima after only a few runs (E.L.A. and A.P.G. data not shown).  However, the next larger dihedral ($D_5$) analogue of $N =$ 137--$N =$ 542--after over one thousand runs appears possibly to be a global minima.  As for the tetrahedral configuration for $N =$ 306, we have no analytic proof of this proposition and given the huge number of 
local minima for an $N$ this large \cite{7} our numerical runs only begin to address the question.  Conversely, we easily found that the next analogue of $N =$ 137, $N =$ 2162, is explicitly not a global energy minimum.  See Figure 3 for dihedral configurations for 
$N =$ 137, 542 and 2162.


Euler's theorem asserts that when a (convex) polyhedron is constructed by joining points 
on a sphere the number of faces ($F$) plus the number of vertices ($V$) is equal to the 
number of edges $(E) + 2; F + V = E + 2$.  For $N> 12$ this has the result that in addition 
to the sixfold coordinated points of a planar two-dimensional lattice (hexamers), there 
must be at least twelve points of fivefold coordination (pentamers).  The tetrahedral 
configurations for $N = 78$ and 306, 
dihedral configurations for $N=137$ and 542, 
and icosadeltahedral configurations for 
$N = 10(h^2 + k^2 + hk) + 2$ have exactly twelve pentamers and the rest hexamers 
(see, e.g., Figure 1$a$, $b$, $c$, $e$,  Figure 2$a$, $b$ and Figure 3$a$, $b$, $c$). Larger analogues ({\it split\/} configurations) of the tetrahedral configuration for $N = 78$ (for $N = $306, 1218, 4886 and for the dihedral configuration for $N =$ 137 ($N =$ 542, 2162) 
(Figure 1$b$, $c$, $e$ and Figure 3$b$, $c$)
are made as follows: in addition to the $N$ charges preexisting place one charge at 
the center of each of the $3N-6$ edges.  (If all the charges were sixfold coordinated 
hexamers there would be $3N$ edges, six must be subtracted from this to take into 
account the twelve fivefold coordinated pentamers.)  Then relax to the final position by a 
local gradient method.  The resulting configuration has $N + (3N-6) = 4N-6 $ charges. 
(See Figure 1$b, c, e$). Some icosadeltahedral configurations can be made ({\it split}) as 
analogues of smaller ones (see e.g. $N = 1242$ in Figure 2$b$, a larger analogue of 
$N = 312$ in Figure 2$a$).  A method for making icosadeltahedral configurations {\it de novo} 
has been discussed previously \cite{12}.  

Conversely, to the cases of $N = 78$ and 306, and $N=$ 137 and 542 for the larger analogues we have studied 
\hbox{($N$ = 1218, 4886, 2162)} we find that adding dislocation defects to the lattice produces 
a configuration with lower energy (Fig.\ 1$c-f$ and Fig.\ 3$c,d$ ).  
Similarly, for $N > 792$ icosadeltahedral 
configurations with dislocation defects, additional fivefold coordinated points and then 
necessary also sevenfold coordinated points (septamers), have lower energy than 
symmetric lattice configurations, while for smaller $N$ the symmetric lattice 
configurations appear to be global minima (see Tables I and II, see Figure 2 for an example), though further numerical testing may show that some such configurations are not global minima.

These numerical results on the $N$ at which tetrahedral, dihedral and icosadeltahedral configurations fail to be global energy minima are in remarkable concordance with a theory given by Dodgson and Moore \cite{13} originally for icosadeltahedral configurations: Using continuum elasticity theory \cite{16} they studied the energy cost of a pair of pentamers, compared with a pure hexagonal lattice, and suggested that dislocation defects--extra fivefold coordinated points, with (necessarily) paired sevenfold coordinated points--would lower lattice strain and energy for $N$ in the $\approx$ 500--1000 range.  Similar reasoning should apply to the tetrahedral configurations in the 
$N = 78$ series and dihedral configurations in the $N=137$ series. 
 Our results given here are strong confirmation of Dodgson and 
Moore’s theory \cite{13}.

The fact that apparently for $N > 792$ all symmetric tetrahedral, dihedral or icosadeltahedral 
lattices are not global minima, along with the exponential growth in good local minima 
may help explain why the number of Mo$_{154}$ anions which self-assemble into a 
spherical superstructure \cite{4} is a non--lattice number 1165, rather than, for example, 
1172 the closest icosadeltahedral lattice or the tetrahedral lattice at 1218, while for small 
$N$ self-assembly often produces a symmetric lattice configuration \cite{2, 3}. 


For icosadeltahedral configurations for $N\le $ 792 whether or not a 
lattice configuration is a potential global minimum depends not 
only on the magnitude of $N$, but also apparently on the details of the
lattice itself.  All lattice $N$ are listed in Tables I and II. As can be
seen for $N$ = 42, 92, 162, 252, 362, 432, 
492, 572, 642, 732
 the icosahedral lattice configuration is
manifestly not the global minimum, while for the other $N$, the lattice configuration appears to be so.  

In an icosadeltahedral lattice $N$ = 10($h^2 + k^2 + hk$) + 2, to go from the center of 
one pentamer to the center of an adjacent pentamer one moves $h$ steps along one 
basis vector fo the lattice, and then $k$ steps in the other. We noted previously 
\cite{12} (also discussed in ref. \cite{17}) that the energy in an icosadeltahedral lattice 
configuration with  a large ratio of $h$ to $k$ ($h\ge k$) may be increased due  to the 
vertices of the pentamers being closely aliged (or perfectly aligned in a lattice with 
$k = 0$).  It has previously been noted that as $N$ grows the icosadeltahedral lattice 
configuration may not be the global minimum \cite{13,14,15,Toomre,Bowick}, 
though we have not seen any explicit published accounting
of the $N$ for which the lattice fails to be a global minimum. This is given in Tables I and 
II (in addition to the rule that for $N > 792$ the lattice is not the global minimum).  
A clear pattern emerges: for an icosadeltahedral configuration with $k$ = 0,
besides the extremely exceptional case of $N$ =12, the icosadeltahedral lattice
configuration is not the global minimum.  For the three small $N=$ 42, 92 and 162 the apparent global minimum  configuration has exactly twelve pentamers, and no dislocation defects, but arranged in an non-icosadeltahedral configuration, likely lowering the energy cost of having the vertices of 
pentamers aligned.  For $N$ larger than 162 in the $k$ = 0 series, the apparent global minimum  incorporates dislocation defects. 
For $N = 432$ and higher for lattices with 
$k = 1$, the lattice is also not the global minimum in accordance with the notion stated 
above that the energy of the lattice is increased by relative alignment of the vertices of 
the pentamers. The data in Tables I and II may be useful in refining theoretical 
predictions for global enery minima: The current theory \cite{16,13} correctly predicts that for $N>1000$ icosadeltahedral and tetrahedral configurations will not be global minima, but does not yet account for various cases for $N<1000$. As well, for $k=0$ other theoretical work \cite{Bowick,17} predicts dislocations lowering the energy for $N>300$, but the first instance
of this is found for $N=252$.

For tetrahedral or dihedral lattices we have not yet been able to find an obvious rule or principle to 
predict for which $N$ the lattice configuration is a global minimum.  Indeed, while all 
groups using a variety of different methods find the $T_h$ configuration a global 
minimum for $N = 78$ \cite{10}, and 
we have found similarly for $N = 306$, there are a number of $N < 100$ for which
$T_h$ configurations are not the global minima \cite{10,11}, and also we have found 
that the next analogue of the global minimum for $N = 100$ which has $T$ (though not 
$T_h$) symmetry, $N = 394$, is not a global minimum (E.L.A. and A.P.G.,
data not shown).  And we see no obvious difference between the $N = 78$ lattice and
 the others to explain why not only for $N = 78$, but the next higher analogue the lattice 
 appears to be the global minimum. 
 Similarly, we have appreciated no obvious reason why 
 the next $D_2$ analogue of 146--578--or 
 the next $D_5$ analogues of
182 and 187--722 and 742--are not global minima, but the next $D_5$ analogue of 137--542--appears to also be a global minimum. 
 The theory of Dodgson and Moore
 \cite{13} does not predict {\it a priori}  that defects would lower a lattice energy for an $N$
  of 306 and 542, 
 as it does  for the next analogues 1218 and 2162. Perhaps, whatever still unknown
reasons that  explains the good minimum for the $T_h$ lattice for $N = 78$ 
and $D_5$ lattice for $N=137$  also
permit the $N = 306$ and 542 analogues to be  global minima.


We thank the anonymous reviewers for comments extremely helpful in revising the 
manuscript and also inspirational leading to finding a new possible global minimum. 
A.P.G. would like to aknowledge  financial support from Spanish MCyT 
under grant No. MAT2003--04887.


\begin{figure}
\epsfxsize=.5\hsize
\begin{center}
\leavevmode
\epsfbox{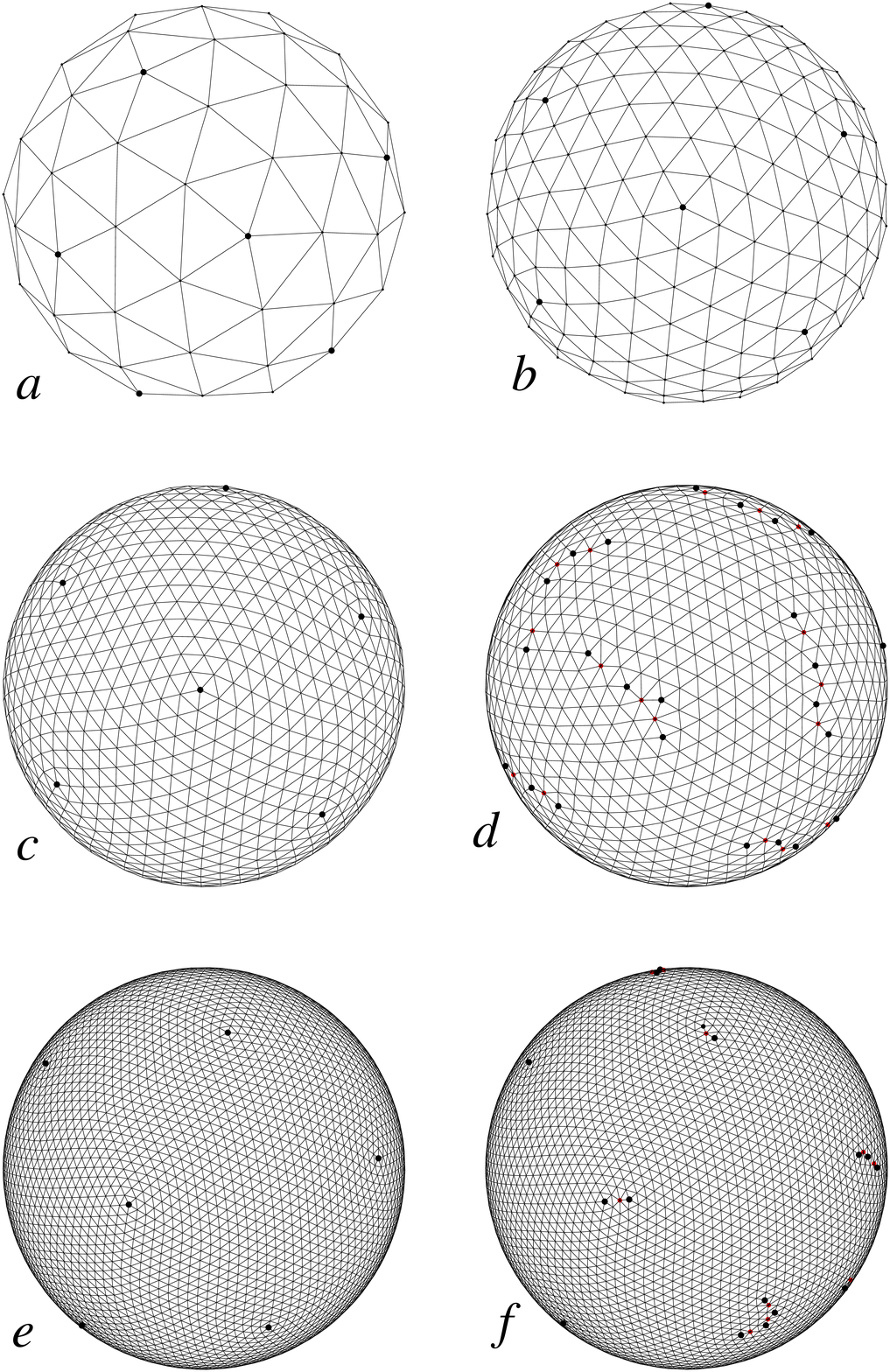}
\end{center}
\caption{Tetrahedral configurations. For $a$) $N$ = 78 
($E$ = 2662.04647), $b$) $N$ = 306 ($E$ =  43862.56978), 
the symmetric tetrahedral lattice configuration appears to be the global 
energy minimum, while for $N$ = 1218 $c$) ($E$ = 718284.03747), $d$) ($E$ = 718281.63110) and 
$N$ = 4886 $e$) ($E$ = 11651484.51295), $f$) ($E$ = 11651440.24177) addition of defects to the lattice produces a configuration of lower energy.  For $N$ = 1218 and 4886 we give the 
lowest energy configuration we have found, though we cannot certain this is a global energy minima.  Fivefold coordinated charges (points) (pentamers) are indicated by large black dots, and sevenfold coordinated charges (septamers) are indicated by small red dots.  The rest of the charges are sixfold coordinated (hexamers). Colors in online version only. }
\end{figure}

\begin{figure}

\epsfxsize=.5\hsize
\begin{center}
\leavevmode
\epsfbox{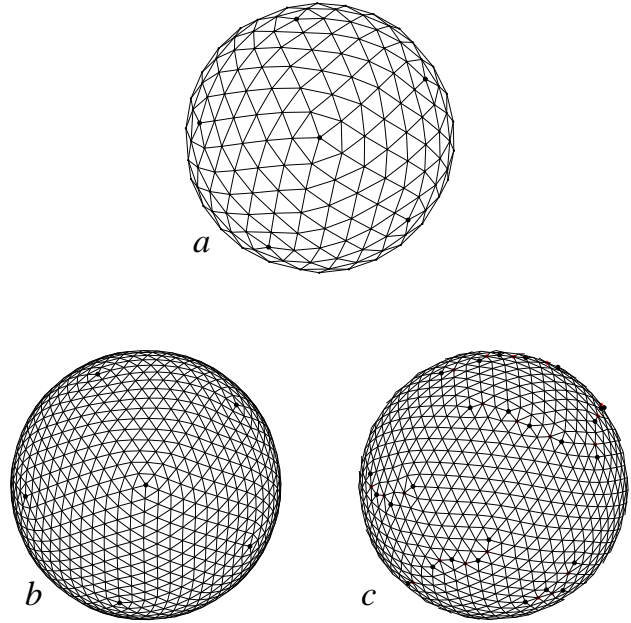}
\end{center}
\caption{Icosadeltahedral configurations. For $a$) $N$ = 312 ($E$ = 45629.36272) the symmetric icosadeltahedral configuration appears to be a global energy minimum, while for $N$ = 1242 $b$) ($E$ = 747107.43183), $c$) ($E$ = 747106.46027) addition of defects produces a configuration of lower energy (though not necessarily the global energy minimum).  As in Figure 1 fivefold coordinated charges (points) (pentamers) are indicated by large black dots, and sevenfold coordinated charges (septamers) are indicated by small red dots.  The rest of the charges are sixfold coordinated hexamers. Colors in online version only.}
\end{figure}

\begin{figure}
\epsfxsize=.5\hsize
\begin{center}
\leavevmode
\epsfbox{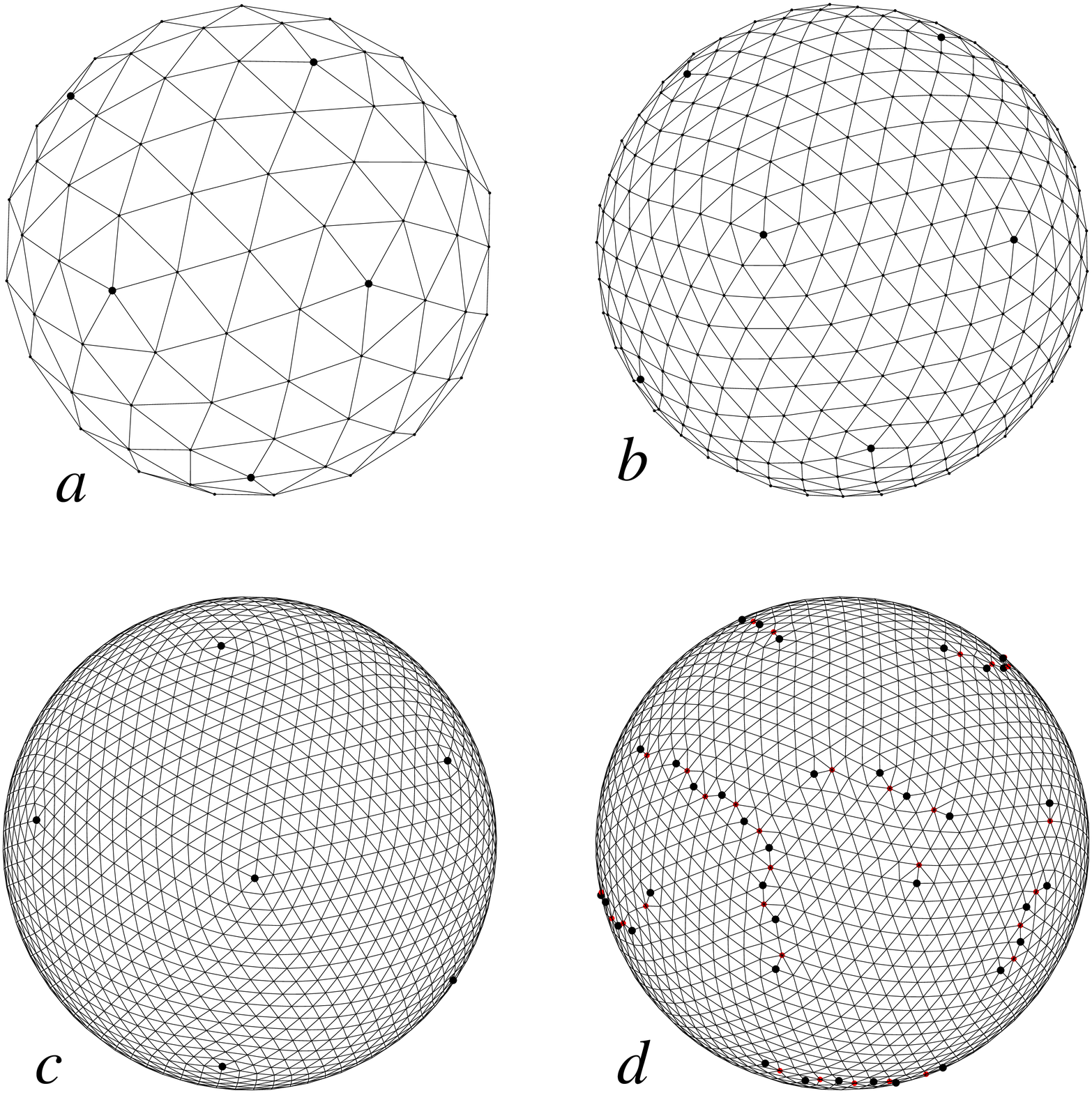}
\end{center}
\caption{Dihedral configurations. For $a$) $N$ = 137 
($E$ =8499.53449), $b$) $N$ = 542 ($E$ =139913.69461), 
the symmetric dihedral lattice configuration appears to be the global 
energy minimum, while for $N$ = 2162 $c$) ($E$ = 2281595.05127), $d$)
 ($E$ = 2281587.48735) addition of defects to the lattice produces a configuration 
 of lower energy.
 We give the lowest energy configuration we have found, though we cannot certain this is
 a global energy minima.  Fivefold coordinated charges (points) (pentamers) are indicated
 by large black dots, and sevenfold coordinated charges (septamers) are indicated by
 small red dots.  The rest of the charges are sixfold coordinated (hexamers). Colors in
  online version only. }
\end{figure}

\newpage

\pagestyle{empty}

\begin{table}
\caption{
Energy of icosadeltahedral configurations with $N<400$.  An ‘*’ indicates a configuration of lower energy, though not necessarily the global minimum. 
For each $N$, here and Table II, we tried fifteen runs--random initial configurations followed by a local gradient descent--to find a configuration with an energy lower than that of the icosadeltahedral lattice.
 For $N = 42$, 92 and 162 the best known configuration has exactly twelve pentamers 
 (and thus no dislocation defects), but does not have icosadeltahedral symmetry.  We 
 note that the “split” larger analogue of the global minimum configuration for $N = 42$ is 
 not the global minimum configuration for $N = 162$, and neither are the split larger 
 analogues of the $N = 92$ and 162 global minima (data not shown). 
}
\begin{tabular}{|lll|}
\hline
Charges & & Energy\\
\hline
          12
 &
 h=           1 k=           0
 &   49.1652530580000     \\
 \hline
          32
 &
 h=           1 k=           1
 &   412.261274651000     \\
 \hline
          42
 &
 h=           2 k=           0
 &   732.256241038000     \\
 &
*Non--icosadeltahedral
 &   732.078107551000     \\
 \hline
          72
 &
 h=           2 k=           1
 &   2255.00119099000     \\
 \hline
          92
 &
 h=           3 k=           0
 &   3745.61873908500     \\
 &
*Non--icosadeltahedral
 &   3745.29163624500     \\
 \hline
         122
 &
 h=           2 k=           2
 &   6698.37449926100     \\
 \hline
         132
 &
 h=           3 k=           1
 &   7875.04534281600     \\
 \hline
         162
 &
 h=           4 k=           0
 &   11984.5514338730     \\
 &
*Non--icosadeltahedral
 &   11984.0503358310     \\
 \hline
         192
 &
 h=           3 k=           2
 &   16963.3383864710     \\
 \hline
         212
 &
 h=           4 k=           1
 &   20768.0530859690     \\
 \hline
         252
 &
 h=           5 k=           0
 &   29544.2821928610     \\
 &
* w/defects
 &   29543.7859569610     \\
 \hline
         272
 &
 h=           3 k=           3
 &   34515.1932926880     \\
 \hline
         282
 &
 h=           4 k=           2
 &   37147.2944184740     \\
 \hline
         312
 &
 h=           5 k=           1
 &   45629.3627238190     \\
 \hline
         362
 &
 h=           6 k=           0
 &   61720.0233978130     \\
 &
* w/defects
 &   61719.3090545160     \\
 \hline
         372
 &
 h=           4 k=           3
 &   65230.0271225660     \\
 \hline

         392
 &
 h=           5 k=           2
 &   72546.2583708950     \\
 \hline

\end{tabular}

\end{table}

\thispagestyle{empty}

\newpage

\thispagestyle{empty}

\begin{table}
\vglue -1cm
\caption{
Energy of icosadeltahedral configurations with $N>400$.  An ‘*’ indicates a configuration of lower energy, though not necessarily the global minimum.  Thus, for $N < 792$ we also cannot be certain that the symmetric icosadeltahedral configurations are the global minima. 
}

\begin{tabular}{|lll|}
\hline
         432
 &
 h=           6 k=           1
 &   88354.2293807250     \\
 &
* w/defects
 &   88354.1906652260     \\
 \hline

         482
 &
 h=           4 k=           4
 &   110318.139920155     \\
 \hline
         492
 &
 h=           7 k=           0
 &   115006.982258289     \\
 &
 *h=           5 k=           3
 &   115005.255889700     \\
 \hline
         522
 &
 h=           6 k=           2
 &   129655.833007858     \\
 \hline
         572
 &
 h=           7 k=           1
 &   156037.879346228     \\
 &
* w/defects
 &   156037.316647696     \\
 \hline
         612
 &
 h=           5 k=           4
 &   178910.494981768     \\
 \hline
         642
 &
 h=           8 k=           0
 &   197100.363816212     \\
 &
* w/defects
 &   197098.637958037     \\
 \hline
         672
 &
 h=           7 k=           2
 &   216171.432658341     \\
 \hline
         732
 &
 h=           8 k=           1
 &   256975.527362500     \\
 &
* w/defects
 &   256974.262894426     \\
 \hline
         752
 &
 h=           5 k=           5
 &   271362.588212841     \\
 \hline
         762
 &
 h=           6 k=           4
 &   278704.548700071     \\
 \hline
         792
 &
 h=           7 k=           3
 &   301321.818305597     \\
 \hline
         812
 &
 h=           9 k=           0
 &   316895.372099956     \\
&
* w/defects
& 316892.668538128\\
 \hline
         842
 &
 h=           8 k=           2
 &   340988.383415978     \\
 &
* w/defects
 &   340987.675098937     \\
 \hline
         912
&
 h=           9 k=           1
 &   400662.383224662     \\
 &
 h=           6 k=           5
 &   400660.132041002     \\

 &
* w/defects
 &   400659.747279004     \\
 \hline
         932
 &
 h=           7 k=           4
 &   418596.898209635     \\
 &
* w/defects
 &   418595.636527970     \\
 \hline
         972
 &
 h=           8 k=           3
 &   455654.618623736     \\
 &
* w/defects
 &   455653.441695822     \\
 \hline
        1082
 &
 h=           6 k=          6
 &   565703.908873765     \\
 &
* w/defects
 &     565703.766602964 \\

 \hline
\end{tabular}
\end{table}

\begin{thebibliography} {99}
\bibitem{1} J.J. Thomson, Philosophical Magazine 7, 237-265 (1904). 
\bibitem{2} D. Caspar, and A. Klug, Cold Spring Harb. Symp. Quant. Biol. 27, 1 (1962). 
\bibitem{3} H.W. Kroto, J.R. Heath, S.C. O’Brien, R.F. Curl, and R.E. Smalley, Nature 318, 162 (1985).
\bibitem{4} T. Liu, E. Diemann, H. Li, A.W. Dress, and A. Muller, Nature 426, 59 (2003).
\bibitem{5} S. Smale, Math. Intelligencer 20 (2), 7 (1998).
\bibitem{6} E.L. Altschuler, T.J. Williams, E.R. Ratner, F. Dowla, and F. Wooten, Phys. Rev. Let. 72, 2671 (1994).
\bibitem{7} T. Erber, and G.M. Hockney, G. M. Phys. Rev. Let. 74, 1482 (1995).
\bibitem{8} A. P\'erez--Garrido, M. Ortu\~no, E. Cuevas and J. Ruiz, J. Phys. A  29, 1973 (1996).
\bibitem{9} J.R. Morris, D.M. Deaven and K.M. Ho, Phys. Rev. B 53, R1740 (1996).
\bibitem{10} T. Erber T. and G.M. Hockney, Adv Chem Phys. 98, 495 (1997). 
\bibitem{11} J.R. Edmundson, Acta Cryst. A49, 648 (1993). 
\bibitem{12} E.L. Altschuler, T.J. Williams, E.R. Ratner, R. Tipton, R. Stong, F. Dowla and F. Wooten, Phys. Rev. Lett. 78, 2681 (1997).
\bibitem{13} M.J.W. Dodgson and M.A. Moore, Rev. B 55, 3816 (1997).  
\bibitem{14} A. P\'erez--Garrido, M.J.W. Dodgson, M.A. Moore, M. Ortu\~no and 
A. D\'\i az--S\'anchez, Phys. Rev. Lett. 79, 1417 (1997). 
\bibitem{15} A. P\'erez--Garrido, M.J.W. Dodgson and M.A. Moore, Phys. Rev. B 56, 3640 (1997). 
\bibitem{Toomre} A. Toomre, unpublished results, discussed in [17,19]
\bibitem{Bowick} M. J. Bowick, D. R. Nelson and  A. Travesset, Phys. Rev. B 62, 8738 
(2000)
\bibitem{16} M.J.W. Dodgson, J. Phys. A 29, 2499 (1996). 
\bibitem{17} M. Bowick, A. Cacciuto, D.R. Nelson, and A. Travesset, Phys. Rev. Lett. 89, 185502 (2002).
\end{thebibliography}
\end{document}